\documentclass{amsart}

\usepackage{amssymb,latexsym,amsfonts,amsmath}

\usepackage{amssymb,latexsym,amsfonts,amsmath}
\usepackage{graphicx,psfrag,epsfig,epsf}
\usepackage{eso-pic}
\usepackage{graphicx}
\usepackage{color}

\usepackage{amsfonts}

\usepackage{subfigure}

\usepackage{amssymb,latexsym,amsfonts,amsmath}
\usepackage{graphicx}

\newtheorem{theorem}{Theorem}[section]

\newtheorem{problem}[theorem]{Problem}

\theoremstyle{definition}

\theoremstyle{remark}

\numberwithin{equation}{section}

\begin{document}

\begin{abstract}                          
In finance industry portfolio construction deals with how to divide the investors' wealth across an asset–-classes' menu in order to maximize the investors' gain. Main approaches in use at the present are based on variations of the classical Markowitz model. However, recent evolutions of the world market showed limitations of this method and motivated many researchers and practitioners to study alternative methodologies to portfolio construction. In this paper we propose one approach to optimal portfolio construction based on recent results on stochastic reachability, which overcome some of the limits of current approaches. Given a sequence of target sets that the investors would like their portfolio to stay within, the optimal portfolio allocation is synthesized in order to maximize the joint probability for the portfolio value to fulfill the target sets requirements. A case study in the US market is given which shows benefits from the proposed methodology in portfolio construction. A comparison with traditional approaches is included. 
\end{abstract}

\title[A stochastic reachability approach to portfolio construction in finance industry]{A stochastic reachability approach \\to portfolio construction in finance industry}
\thanks{This work has been partially supported by the Center of Excellence for Research DEWS, University of L'Aquila, Italy and by the National Science Foundation CAREER award 0717188.}

\author[Giordano Pola and Gianni Pola]{
Giordano Pola$^{1}$ and Gianni Pola$^{2}$}
\address{$^{1}$
Department of Electrical and Information Engineering, Center of Excellence DEWS,
University of L{'}Aquila, Poggio di Roio, 67040 L{'}Aquila, Italy}
\email{giordano.pola@univaq.it}
\urladdr{
http://www.diel.univaq.it/people/pola/
}

\address{$^{2}$Quantitative Research Department, Cr\'edit Agricole Asset Management SGR, Piazza Missori 2, 20122 Milan, Italy}
\email{gianni.pola@caam.com}

\maketitle

\section{Introduction}
In finance industry portfolio construction deals with how to divide the investor's wealth across some asset--classes selected from a
given asset--classes' menu in order to maximize the investor's gain. An asset--class is a specific category of investments such as stocks, bonds, cash, and commodities. In quantitative finance portfolio construction is achieved by an optimization process. A pioneering work in this regard is the well--known Markowitz model \cite{markowitz}. Given a target performance, Markowitz method provides
the optimal strategy which minimizes the investment risk. This method and variations of it, are extensively used at the present by many asset managers. However, recent evolutions of the world market showed limitations of this method and motivated many researchers and practitioners to study alternative methodologies to portfolio construction.
Main drawbacks of the Markowitz method are summerized in:

\begin{itemize}
\item The asset--classes' performance is assumed to be multivariate gaussian distributed.
\item Investors make a one--shot allocation; the model does not face the portfolio re--balancing during the
investment life--time.
\end{itemize}

The first drawback is shown by basic econometric analysis of the world market to be inappropriate to capture relevant statistical properties of asset--classes' performances, while the second one does not allow the investor to have a tight control of portfolio evolution during the investment life--time.
Recently, Optimal Dynamic Asset Allocation (ODAA) proposes a methodology to overcome such limitations. ODAA deals with how to optimally allocate a
multi--period investment. First studies in ODAA faced the problem on how to divide the investment among stock
and money markets (see e.g. \cite{MER} and
\cite{SAM}). More recently, some work appeared in the literature
concerning optimal strategies for long--lived investors under stochastic
investment opportunities (see e.g. \cite{BREXIA} and the references therein). In
particular, the work in \cite{BRE} studies portfolio re--balancing in the presence of
stochastic variation in the interest rate, the work in \cite{BREXIA} considers the effects
of inflation in a portfolio of stock or nominal bonds, the work in \cite{BAR} and \cite{XIA} take into account the uncertainty in the asset returns prediction. \\
In this paper we consider the ODAA problem and we propose a methodology which is based on
recent results on stochastic reachability, see e.g.  \cite{Pola06,AbateAut08,BujorianuHSCC04}. Stochastic reachability deals with the synthesis of a control strategy which maximizes the probability for the state of a
stochastic dynamical control system to be in (or to reach) a desired set, representing the set of ``good states''.
In this paper we reformulate ODAA as an appropriate stochastic reachability optimal control problem. Given a specified finite time horizon and a sequence of target sets that the investors would like their portfolio to stay within, the optimal portfolio allocation is synthesized in order to maximize the joint probability for the portfolio value to fulfill the target sets requirements.
Main features of this formulation of the ODAA problem are summerized in:
\begin{itemize}
\item[(F1)] No specific probability distribution assumed on the asset--classes' performances.
\item[(F2)] Sensitivity to market movements.
\item[(F3)] Portfolio evolution control during its life-time.
\end{itemize}
We stress that the above features overcome Markowitz limits to portfolio construction. 
The benefits from the proposed approach are illustrated by means of a case study in the US market. 
A comparison with the solution provided by traditional methodologies based on Markowitz optimizer is also included. 
A preliminary investigation on the formalization of the ODAA problem in terms of a stochastic reachability problem appeared in the conference
publication \cite{PolaCDC06}. The present paper provides a detailed and mature description of the results announced in \cite{PolaCDC06}, including econometric analysis of the world market and a case study in the US market.\\
This paper is organized as follows. In Section \ref{sec2} traditional approaches to portfolio construction are briefly recalled and discussed. In Section \ref{sec3} an econometric analysis of the main world market is presented, which shows limitations of traditional approach to portfolio construction. Section \ref{sec4} is devoted to the novel approach on portfolio construction that we propose. Benefits from this approach are illustrated by means of a case study in the US market in Section \ref{sec5}. Details on economic indexes used in such case study are reported in the Appendix. Finally Section \ref{sec6} offers some conclusive remarks and outlook.

\section{An Introduction to Portfolio Construction in Finance Industry}\label{sec2}

\subsection{Financial Asset--Classes}
Portfolio construction deals with how to divide the investor wealth across some asset--classes' in order to maximize the investor gain.
An asset--class is a specific category of investments such as bonds, stocks, and commodities. Assets within the same
class generally exhibit similar risk characteristics, behave similarly in the market--place, and are subject to the same laws and regulations.
Due to unpredictability of their behaviour, asset--classes dynamics is usually modeled by means of stochastic processes. 
Consider an investment universe made of $m$ asset--classes. Given
$k\in \mathbb{N}$ define the random vector:
\[
w_{k}=\left[
\begin{array}
[c]{cccc}%
w_{k}(1) & w_{k}(2) & \cdots & w_{k}(m)
\end{array}
\right]  ^{T}\in \mathbb{R}^{m},
\]
whose entries are the performances or equivalently the returns, which are associated with the asset--classes at time $k$. The performance $w_{k}(i)$ 
is defined as the percentage variation of the asset--class price $z_{k}(i)$ in the time interval $[k-1,k]$, i.e.
\begin{equation}
w_{k}(i)=\frac{z_{k}(i)-z_{k-1}(i)}{z_{k-1}(i)}, \label{percRET}%
\end{equation}
where $z_{k}(i)$ and $z_{k-1}(i)$ correspond to the prices of the $i$--th asset--class at times $k$ and $k-1$,
respectively. 
We assume that $w_{k}$ is a correlated random vector. 
%
%
%
The \emph{Expected Return} (ER) and \emph{Standard Deviation} (SD) for the $i$--th asset--class at time $k$ are
defined respectively by:
\[
\mu_{k}(i)= \mathbb{E}[w_{k}(i)],\hspace{5mm} \sigma_{k}(i)=( \mathbb{E}[(w_{k}(i)-\mu_{k}(i))^{2}])^{1/2},
\]
where $ \mathbb{E}[\,\cdot\,]$ is the usual expectation operator. We denote by $\mu_{k}\in \mathbb{R}^{m}$ and $\sigma_{k}\in \mathbb{R}^{m}$ the collection of the asset--classes' ERs and SDs at time $k$, i.e.%
\[
\begin{array}
[l]{ll}
\mu_{k}=\left[
\begin{array}
[c]{cccc}%
\mu_{k}(1) & \mu_{k}(2) & \cdots & \mu_{k}(m)
\end{array}
\right]^{T}  \in \mathbb{R}^{m},\nonumber\\
\sigma_{k}=\left[
\begin{array}
[c]{cccc}%
\sigma_{k}(1) & \sigma_{k}(2) & \cdots & \sigma_{k}(m)
\end{array}
\right]^{T}  \in \mathbb{R}^{m}.\nonumber
\end{array}
\]
%
Asset--classes' standard deviation is known in finance--industry as \textit{volatility} and it is a risk measure expressing the variability of the
asset--class performance around the ER.
Covariance matrix (CM) at time $k$ is defined by:
\[
\Sigma_{k}(i,j)= \mathbb{E}[(w_{k}(i)- \mathbb{E}[w_{k}(i)])(w_{k}(j)- \mathbb{E}[w_{k}(j)])].%
\]

\begin{figure}[ptb]
\begin{center}
\includegraphics[scale=0.4]{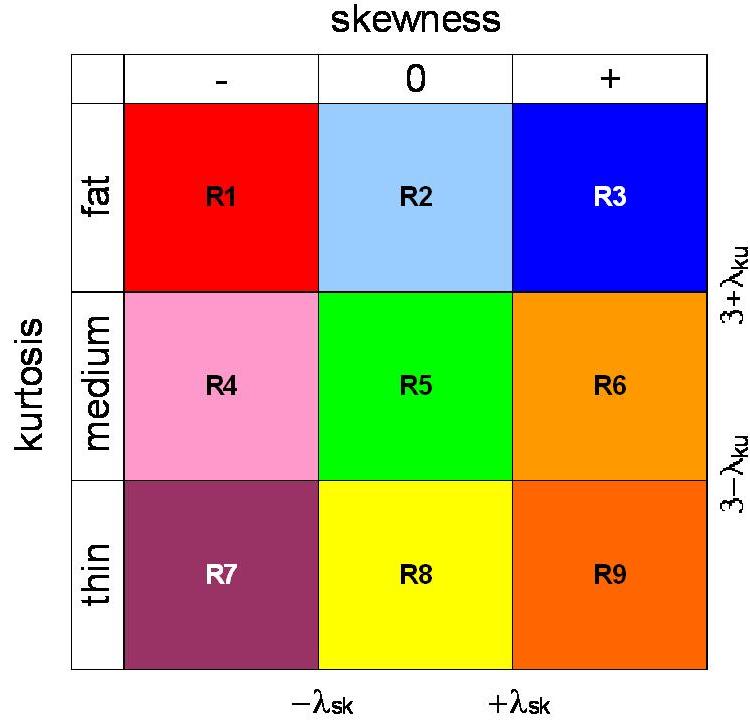}
\end{center}
\caption{\emph{Skewness--Kurtosis Plane}.}
\label{fig33}%
\end{figure}

Quantities ER, SD and CM relate to the first and second moments associated with the random vector $w_{k}$. Univariate higher order statistical indicators of interest are \textit{Skewness} (SK) and \textit{Kurtosis} (KU) which are defined respectively by:
\[
\eta_{k}(i) = \mathbb{E}\left[\left(\frac{w_{k}(i)-\mu_{k}(i)}{\sigma_{k}(i)}\right)^{3}\right], \hspace{5mm}
\theta_{k}(i) = \mathbb{E}\left[\left(\frac{w_{k}(i)-\mu_{k}(i)}{\sigma_{k}(i)}\right)^{4}\right], 
\]
and related to the third and fourth moments, respectively. Skewness provides a measure of asymmetry in the distribution. Kurtosis quantifies the occurrence of rare events far away from the ER. Gaussian--distributed random variable are characterized by a skewness value equal to $0$ and by a kurtosis value equal to $3$; higher values of KU indicate asset-classes with an higher probability of extreme--events, while lower values indicate that returns are more predictable and clustered around the ER. Significant deviations from gaussianity are usually assessed throughout statistical tests.
Given a confidence level $CL$, which quantifies the accuracy in the statistical test, and a data--sample size $N=250$, corresponding to the working days per year, an asset--class is said to be:
\begin{itemize}
  \item \textit{Positive Skewed}, if the sample skewness $SK\geq \lambda_{sk}$;
  \item \textit{Gaussian-like Skewed}, if the sample skewness $SK\in ]\lambda_{sk},\lambda_{sk}[$;
  \item \textit{Negative Skewed}, if the sample skewness $SK\leq -\lambda_{sk}$,
\end{itemize}
where: 
\[
\lambda_{sk}=\left(\frac{6\,CL}{-1+N}\right)^\frac{1}{2}.
\]
Moreover an asset--class is said to be:
\begin{itemize}
  \item \textit{Leptokurtic}, if the sample kurtosis $KU\geq 3+\lambda_{ku}$;
  \item \textit{Mesokurtic}, if the sample kurtosis $KU\in ]3-\lambda_{ku},3+\lambda_{ku}[$;
  \item \textit{Platykurtic}, if the sample kurtosis $KU\leq 3-\lambda_{ku}$,
\end{itemize}
where: 
\[
\lambda_{ku}=\left(\frac{24\,CL}{-1+N}\right)^\frac{1}{2}.
\]
Each asset--class skewness--kurtosis pair can be represented in the skewness--kurtosis plane. From the above taxonomy the skewness--kurtosis plane can be partitioned into nine regions of interest, as depicted in Figure \ref{fig33}: 

\begin{itemize}

\item \textit{Region 1}: negative skewed and leptokurtic;

\item \textit{Region 2}: gaussian--like skewed and leptokurtic;

\item \textit{Region 3}: positive skewed and leptokurtic;

\item \textit{Region 4}: negative skewed and mesokurtic;

\item \textit{Region 5}: gaussian--like skewed and mesokurtic;

\item \textit{Region 6}: positive skewed and mesokurtic;

\item \textit{Region 7}: negative skewed and platykurtic;

\item \textit{Region 8}: gaussian--like skewed and platykurtic;

\item \textit{Region 9}: positive skewed and platykurtic.

%
%
%
%
%
%

\end{itemize}

Each of the regions present specific statistical features. For example, Region 1 is characterized by an excess of negative\footnote{Negative and positive events are evaluated with respect to the ER, i.e. an event is negative/positive when its evaluation is less/bigger than the ER.} rare events, Region 3 by an excess of positive$^{1}$ rare events, while Region 5 is not characterized by neither a remarkable asymmetry nor by a remarkable fat or thin kurtosis; asset--classes whose skewness kurtosis pair are in Region 5 can be assumed to be gaussian distributed.
%

\subsection{Traditional Approaches to Portfolio Construction in Finance Industry}

An asset allocation is a vector $u\in \mathbb{R}^{m}$ where the $i$--th
entry expresses the amount of the investment in the $i$--th asset--class. For
example $u=[0.30\hspace{2mm}0.50\hspace{2mm}0.20]$ indicates that
we are investing $30\%$ of our wealth in the first asset--class, $50\%$ in the
second, and $20\%$ in the latter. Let $u_{k}$\ be the asset allocation at
time $k\in \mathbb{N}$. Some constraints are usually imposed on $u_{k}$ in the
investment process. The most common ones in finance industry are:

\begin{itemize}
\item \emph{Budget constraint:} $%
{\textstyle\sum_{i=1}^{m}}
u_{k}(i)=1$. This equality means that the investor's wealth is fully invested
in the portfolio;

\item \emph{Long--only constraint:} $u_{k}(i)\geq0$, for any $i=1,\ldots,m$. This
inequality implies that short selling is not allowed;

\item \emph{Risk--budget constraint:} $(u_{k-1}^{T}\Sigma_{k}u_{k-1})^{1/2}%
\leq\sigma_{max}$. This inequality indicates that portfolio risk is
up--ward bounded by $\sigma_{max}$.
\end{itemize}

We denote by $U_{k}$ the collection of all asset--allocations allowed in the investment process at time $k\in\mathbb{N}$.
Portfolio performance or equivalently portfolio return $r_{k}$, can be defined as done for a single asset--class in equation (\ref{percRET}), resulting in:%
\begin{equation}
r_{k}=\frac{x_{k}-x_{k-1}}{x_{k-1}}, \label{eq1}%
\end{equation}
where $x_{k}$ and $x_{k-1}$ are the portfolio value at times
$k$ and $k-1$, respectively. It can be shown \cite{luenberger} that the portfolio
performance $r_{k}$ at time $k$ is given by:
\begin{equation}
r_{k}=u_{k-1}^{T}w_{k}. \label{eq2}%
\end{equation}
The portfolio ER at time $k$ is therefore given by:%
\[
\tilde{\mu}_{k}= \mathbb{E}[r_{k}]=u_{k-1}^{T} \mu_{k}.
\]
Analogously the portfolio SD at time $k$ can be expressed by:
\[
\tilde{\sigma}_{k}=( \mathbb{E}[(r_{k}-\mu_{k})^{2}])^{1/2}=(u_{k-1}^{T}%
\Sigma_{k}u_{k-1})^{1/2}.
\]


Traditional approaches to portfolio construction in finance industry make use of (variations of) the classical Markowitz method \cite{markowitz}.
Markowitz method provides an optimal solution to the asset--classes'
allocation problem.
The optimality criterion considered is the \textit{risk minimization}.
Let us consider an asset--classes' menu, characterized by ERs $\mu_{k}$, CM $\Sigma_{k}$ and a target return $\overline{r}$.
Investors in Markowitz world wish to minimize portfolio risk among the portfolios ensemble with target return $\overline{r}$. The Markowitz optimization problem is therefore formalized as follows:
\begin{equation}
\left\{
\begin{array}
[c]{l}%
\min_{u_{k}\in U_{k}}u_{k}^{T}\Sigma_{k}u_{k},\\
u_{k}^{T}\mu_{k}=\overline{r}.
\end{array}
\right.
\label{opt_mark}%
\end{equation}
Feasible target returns $\bar{r}$ belong to the interval $[\overline{r}_{min},\overline{r}_{max}]$, where:
\begin{eqnarray}
\overline{r}_{min} =\big(\arg\inf_{u_{k}\in U_{k}}u_{k}^{T}\Sigma_{k}%
u_{k}\big)^{T}\mu_{k}, &
\overline{r}_{max} =\sup_{u_{k}\in U_{k}}u_{k}^{T}\mu_{k}.\nonumber
\end{eqnarray}
By performing the optimization problem in (\ref{opt_mark}), with target return $\overline{r}$ ranging in $[\overline{r}_{min},\overline
{r}_{max}]$, we obtain the collection of all optimal portfolios, known in the literature as the
\emph{Efficient Frontier} \cite{luenberger}. Efficient algorithms are known in the literature for solving the optimization problem in (\ref{opt_mark}). While being handable for practical computation, Markowitz method exhibits some drawbacks, summarized in:

\begin{itemize}
\item The asset--classes' performance is assumed to be multivariate gaussian distributed: the only risk--figures involved in the portfolio optimization are contained in the covariance matrix CM. 
\item Investors make a one--shot allocation, the model does not face the problem of how (optimally) rebalance portfolio during the
investment life--time.
\end{itemize}

In the next section we will show the impact of such drawbacks in portfolio construction.
%


\section{Limits of Gaussian Models in Portfolio Construction}\label{sec3}
\begin{figure}[ptb]
\begin{center}
\includegraphics[scale=0.20]{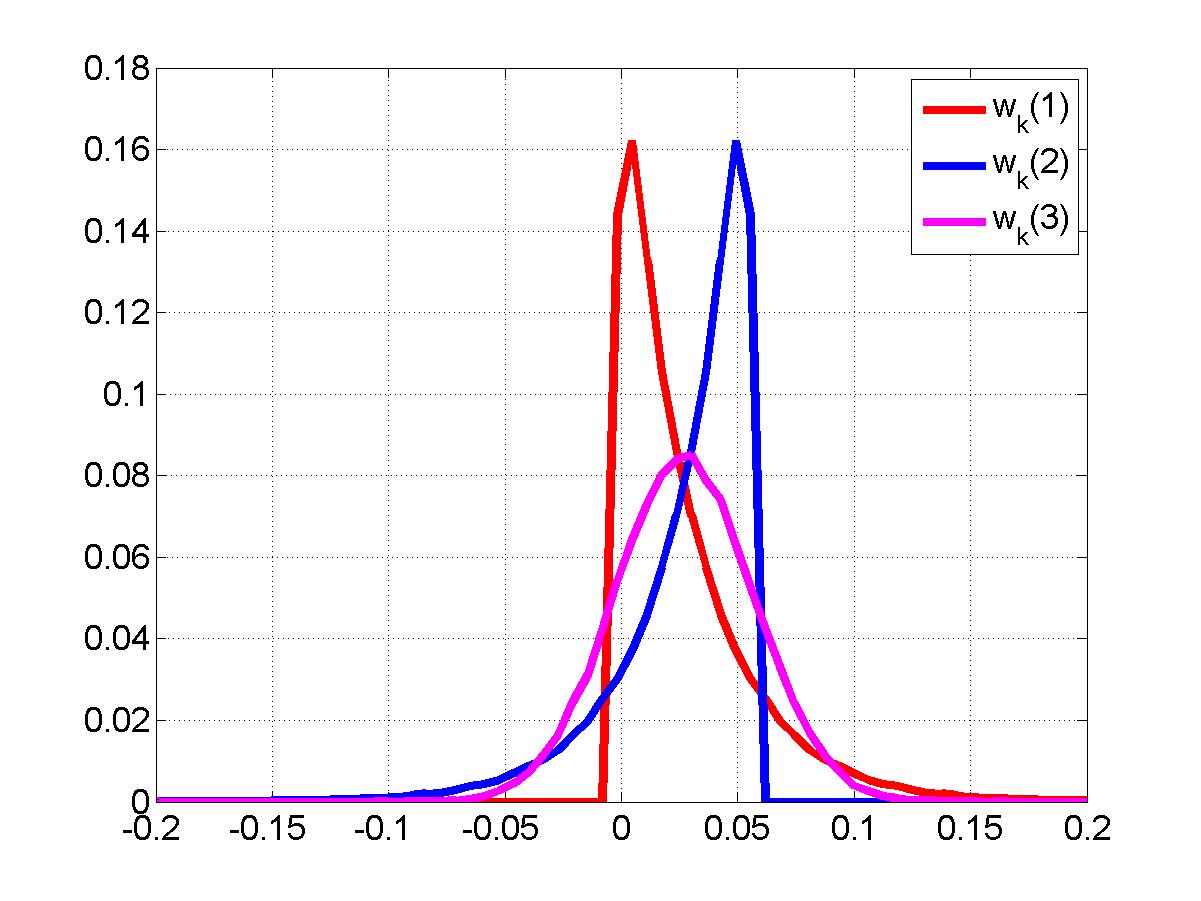}
\end{center}
\caption{\emph{Synthetic data--set}. Marginal probability density functions.
}%
\label{fig2}%
\end{figure}

\begin{figure}[ht]
\centering
\subfigure[JPY-EUR in the skewness--kurtosis plane.]{
\includegraphics[scale=0.185]{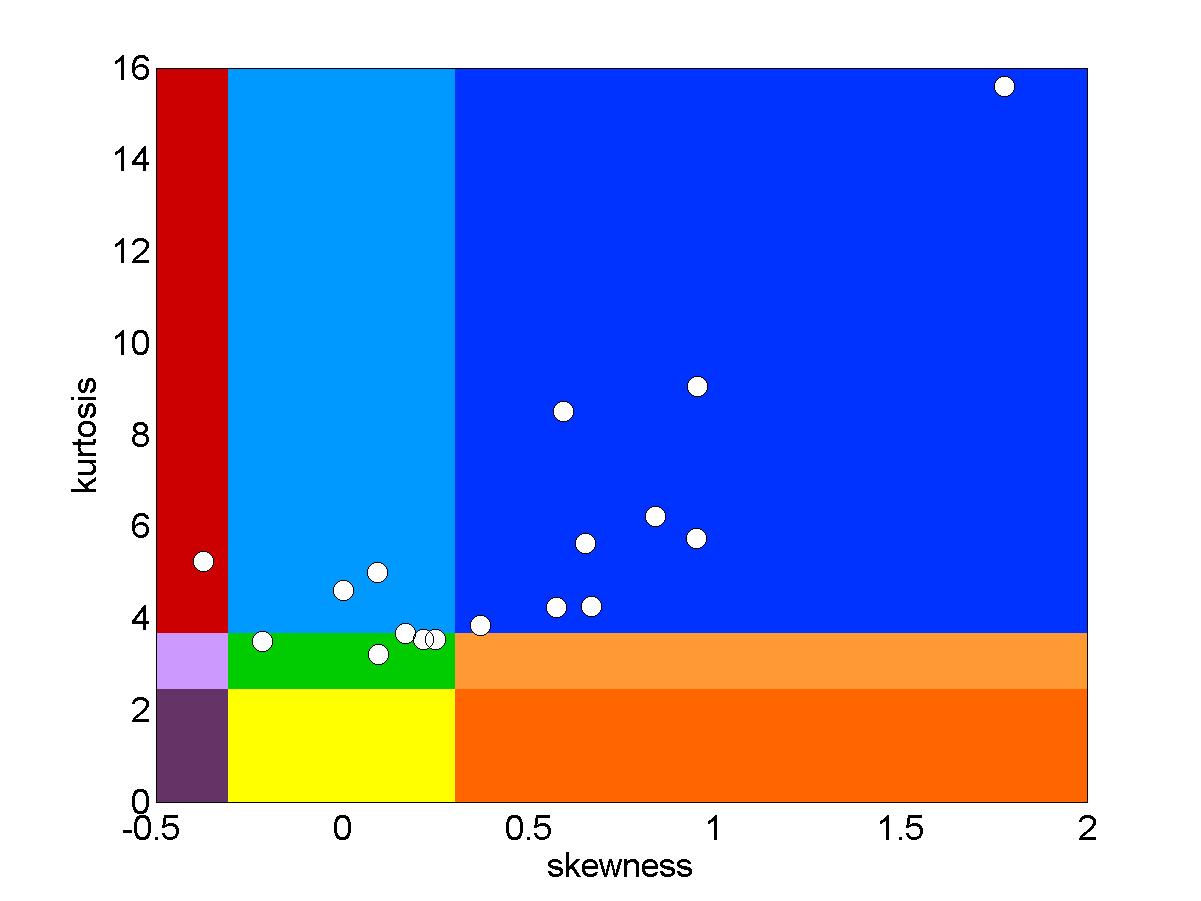}
\label{fig5}
}
\subfigure[Emerging Market Govy in the skewness--kurtosis plane.]{
\includegraphics[scale=0.185]{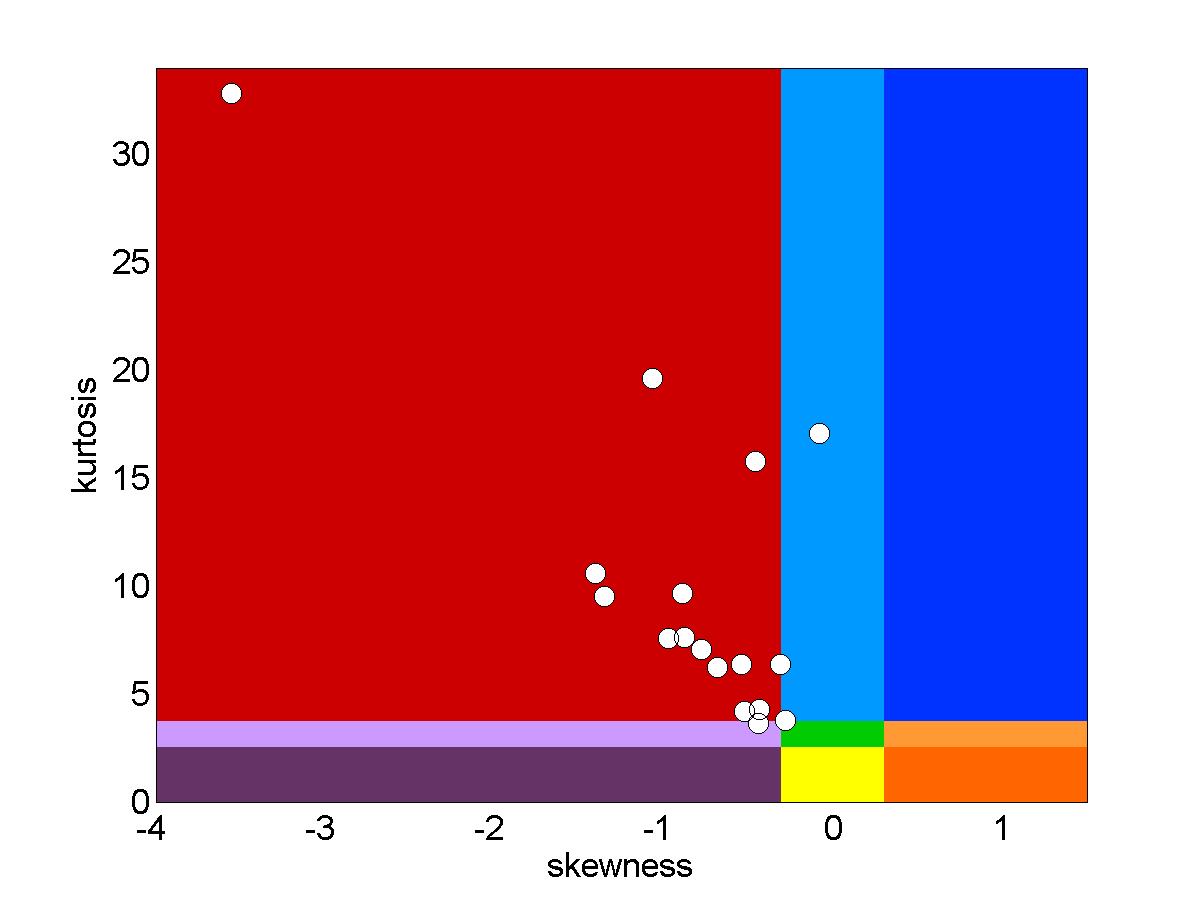}
\label{fig4}
}
\label{fig:subfigureExample}
\caption[]{JPY-EUR and Emerging Market Govy in the skewness--kurtosis plane. Each dot represents a skewness--kurtosis pair in each year ranging from 1992 to 2008.}
\end{figure}

\begin{figure}[ptb]
\begin{center}
\includegraphics[scale=0.83]{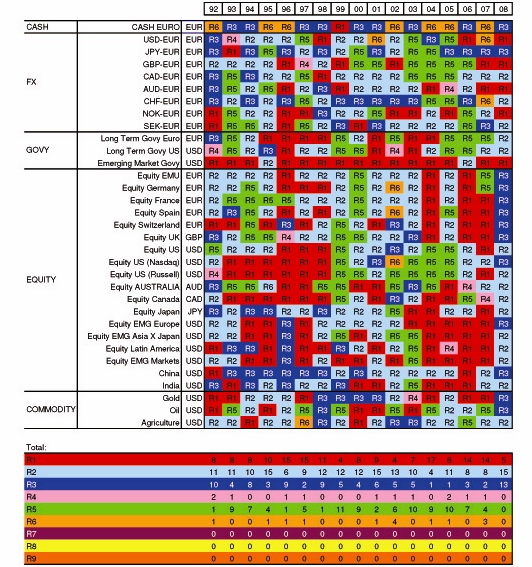}
\end{center}
\caption[above]{\emph{Econometric analysis of the main world market.}}
\label{fig12}%
\end{figure}

We start by providing an academic example which illustrates implications of gaussian--based asset--classes modeling in portfolio construction.
Consider an asset--classes' ensemble composed of three un--correlated synthetic asset--classes.
The asset--classes returns are modeled as follows:
\begin{equation}
\begin{array}
[c]{ccc}%
w_{k}(1)=\gamma, & w_{k}(2)=2\rho-\gamma, & w_{k}(3)=\eta,
\end{array}
\label{mpdf}
\end{equation}
where $\rho\in \mathbb{R}$, random variable $\gamma$ is $\Gamma(\alpha,\beta)$ distributed with
$\alpha=1$ and $\beta=\rho$, and $\eta$ is a gaussian random variable with
ER and SD equal to $\rho$. In this example we set $\rho=0.03$. Figure \ref{fig2} illustrates the marginal probability density functions of the random variables in (\ref{mpdf}). \\
It is possible to show that the considered asset--classes share the same ER and SD, although they are very different in terms of skewness and kurtosis. The following table reports statistical indicators associated with $w_{k}(1)$, $w_{k}(2)$ and $w_{k}(3)$:

\begin{equation}
\begin{tabular}
[c]{lrrrr}\hline
Asset & ER & SD & SK & KU\\\hline
Asset 1 & 0.03 & 0.03 & 2 & 9\\
Asset 2 & 0.03 & 0.03 & -2 & 9\\
Asset 3 & 0.03 & 0.03 & 0 & 3\\\hline
\end{tabular}
\label{TabExample1}
\end{equation}
With reference to the skewness--kurtosis plane in Figure \ref{fig33}, Asset $1$ is in Region $3$, Asset $2$ is in Region $1$ and Asset $3$ in Region $5$.
Since ER and SD of the asset--classes are the same, asset allocation based on these statistical indicators is not able to distinguish among asset--classes. Indeed, Markowitz optimizer selects the optimal solution $u_{0}^{\ast}$ of (\ref{opt_mark}) given by the equally--weighted portfolio, 
i.e.
\[
u_{0}^{\ast}=[%
\begin{array}
[c]{ccc}%
1/3 & 1/3 & 1/3
\end{array}
]^{T}.
\]
On the other hand a simple inspection of the table in (\ref{TabExample1}) indicates that the three asset--classes present important
differences in the marginal probability distribution. A prudent investor would invest in Asset $1$ which is characterized by low probability to suffer negative returns but low probability to reach large positive returns. On the way around Asset $2$ is much riskier than Asset $1$ but it guarantees higher probability to get large positive returns.
This simple example shows limitations of portfolio construction based on gaussian modeling of asset--classes. 
On the other hand deviations from gaussianity are evident in the main world market asset--classes. \\
In the following we report a basic econometric analysis of the main world markets. We analyze $33$ indices from January $1$--st $1992$ to December $31$--th $2008$. Indices are in local currencies (see Appendix A) and daily--based (closing price values). In order to test the gaussian hypothesis we perform the Jarque--Bera test to $95\%$ Confidence Level on the indices' returns. We considered $1$--year non--overlapping windows from $1992$ to $2008$; this choice allows us to monitor the time--evolution of higher order statistics year by year.\\
Results are shown in Figure \ref{fig12}. Asset--classes in Region R5, for which gaussinity hypothesis can be assumed account at a maximum of $33\%$ in 1999 and a minimum of $0\%$ in 2008. Asset--classes in Regions R1 and R3 account at a maximum of $72\%$ in 1996 and a minimum of $27\%$ in 1999, 2002 and 2005. Some meaningful specific cases are reported below:
\begin{itemize}

\item \textit{JPY-EUR}. Figure \ref{fig5} depicts statistical historical data associated with JPY-EUR in the skewness--kurtosis plane. Each dot represent a skewness--kurtosis pair in each year ranging from 1992 to 2008. Only $5$ over $17$ dots are located in Region 5;

\item \textit{Emerging Market Govy}. Figure \ref{fig4} depicts statistical historical data associated with Emerging Market Govy in the skewness--kurtosis plane. Each dot represent a skewness--kurtosis pair in each year ranging from 1992 to 2008. Only $1$ over $17$ dots is located in Region 5.

\end{itemize}

%
%

This econometric analysis highlights the evidence and the impact of higher order statistics in common--used risk measure.

\section{A Novel Approach to Portfolio Construction}\label{sec4}

In this section we illustrate an alternative approach to portfolio construction, based on stochastic optimal control. A stochastic dynamical control system modeling the portfolio value dynamics can be derived by combining equations in
(\ref{eq1}) and (\ref{eq2}), resulting in:%
\begin{equation}
\label{system}
\begin{array}
[c]{ll}
x_{k+1}=x_{k}(1+u_{k}^{T}w_{k+1}),& k\in \mathbb{N},\\
\end{array}
\end{equation}
where:

\begin{itemize}
\item $x_{k}\in X= \mathbb{R}$ is the state, representing the portfolio value at time $k$;

\item $u_{k}\in U_{k}\subseteq \mathbb{R}^{m}$ is the control input, representing the portfolio allocation at time $k$;

\item $w_{k}\in \mathbb{R}^{m}$ is a random vector describing the
asset--classes' returns at time $k\in\mathbb{N}$.

\end{itemize}

Let $(\Omega,\mathcal{F},P)$ be the probability space associated with the stochastic system in (\ref{system}). Portfolio value $x_{k}$ at time $k=0$ is assumed to be known and set to $x_{0}=1$. The mathematical model in (\ref{system}) is characterized by:

\begin{itemize}
\item[(F1)] No specific distribution on the asset--classes' returns.
\end{itemize}
As already discussed in the previous section this feature is important in asset--allocation. We model asset--classes' returns by means of Mixture of Multivariate Gaussian Models (MMGMs), which provide an accurate modeling of non--gaussian distributions while being handable for practical implementations. 
We recall that
a random vector $Y$ is said to be distributed according to a MMGM if its probability density function $p_{Y}$ can be expressed as the convex combination of probability density functions $p_{Y_{i}}$ of some multivariate gaussian random variables $Y_{i}$, i.e.
\[
p_{Y}(y)=\sum_{i=1}^{N}\lambda_{i}p_{Y_{i}}(y),
\hspace{5mm} \lambda_{i}\in [0,1],
\hspace{5mm} \sum_{i=1}^{N} \lambda_{i}=1.
\]
Some further constraints are usually imposed on coefficients $\lambda_{i}$ so that the resulting random variable $Y$ is well behaved, by requiring for example semi--definiteness of covariance matrix and unimodality in the distribution.
The interested reader can refer to \cite{IAN} for a comprehensive exposition of main properties of MMGMs. \\
The class of control inputs that we consider in this paper is the one of Markov policies \cite{bertsekas}. Given a finite time horizon $N\in\mathbb{N}$ a Markov policy is defined by the sequence
\begin{equation}
\pi=\{u_{0},u_{2},...,u_{N-1}\}
\nonumber
\end{equation}
of measurable maps $u_{k}:X\rightarrow U_{k}$. Denote by $\mathcal{U}_{k}$ the set of measurable maps $u_{k}:X\rightarrow U_{k}$
and by $\Pi_{N}$ the collection of Markov policies. For further purposes let be $\pi ^{k}=\left\{u_{k},u_{k+1},...,u_{N-1}\right\}$.
Markov policies adequately model portfolio allocations, since they allow the investor to have:
\begin{itemize}
\item[(F2)] Sensitivity to market movements.
\end{itemize}

Indeed, control input $u_{k}$ depends on $x_{k}$ and hence on the market performance at time \mbox{$t\in [k-1,k]$}.\\
%
%
Let us consider a finite time horizon $N$ which represents the life--time of the considered investment.
Our approach in the portfolio construction deals with how to select a Markov policy $\pi$ in order to fulfill some specifications on the portfolio value $x_{k}$ at times $k=1,\ldots,N$. The specifications are defined by means of a sequence of target sets $\{X_{1},X_{2},...,X_{N}\}$ with $X_{i}\subseteq X$. The investor wishes to have a portfolio value $x_{k}$ at time $k$ that is in $ X_{k}$. Typical target sets $X_{k}$ are of the form $X_{k}=[\underbar{x}_{k},+\infty[$ and ask for achieving a performance that is downside bounded by $\underbar{x}_{k}\in \mathbb{R}$. An example of target sets pattern is illustrated in equation (\ref{sigmas}) and models an investor specification which requires to beat a target return $7\%$ (annualized value) at maturity at the end of the investment life--time.
This formulation of specifications allows the investor to have a
\begin{itemize}
\item[(F3)] Portfolio evolution control during its life-time,
\end{itemize}
since target sets $ X_{k}$ depend on time $k$. 
The portfolio construction problem is then formalized as follows:

\begin{problem}
\textit{(Optimal Dynamic Asset Allocation (ODAA))} Given a finite time horizon
$N\in \mathbb{N}$ and a sequence of target sets
\begin{equation}
\{X_{1},X_{2},...,X_{N}\},
\label{targetsets}
\end{equation}
where $X_{k}$ are Borel subsets of $X$, find the optimal Markov policy $\pi$
that maximizes the joint probability quantity
\begin{equation}
P(\{\omega \in \Omega: x_{0}\in X_{0},x_{1}\in X_{1},\ldots,x_{N}\in X_{N}\}). \label{OptProb}%
\end{equation}
\label{ODAApb}
\end{problem}

The ODAA Problem can be solved by using a dynamic programming approach \cite{bertsekas} and in particular by resorting to recent results on stochastic reachability (see e.g. \cite{Pola06,AbateAut08,AbatePHDThesis,BujHSCC03,BujorianuHSCC04}). 
Since the solution of Problem \ref{ODAApb} can be obtained by a direct application of the results in the work of \cite{Pola06,AbateAut08} in the following we only
report the basic facts which lead to the synthesis of the optimal portfolio allocation. 
Given $x\in X$ and $u\in \mathbb{R}^{m}$ denote by $p_{f(x,u,w_{k})}$ the probability density function of random variable:
\begin{equation*}
f(x,u,w_{k+1})=x(1+u^{T}w_{k+1}),
\end{equation*}
associated with the dynamics of the system in (\ref{system}).
Given the sequence of target sets in (\ref{targetsets}) and a Markov policy $\pi$ we introduce the following cost
function $V$ which associates a real number $V(k,x,\pi ^{k})\in \lbrack 0,1]$
to a triple $(k,x,\pi ^{k})$ by:
\begin{equation*}
V(k,x,\pi ^{k})=\left\{
\begin{array}{ll}
I_{X_{k}}(x), & \text{if }k=N, \\
\int_{X_{k+1}}V(k+1,z,\pi ^{k+1})p_{f(x,u_{k},w_{k+1})}(z)dz, & \text{if }k=N-1,N-2,...,1,0,%
\end{array}%
\right.
\end{equation*}%
where $I_{X_{N}}(x)$ is the indicator function of the Borel set $X_{N}$, i.e. $I_{X_{N}}(x)=1$ if $x\in X_{N}$ and $I_{X_{N}}(x)=0$, otherwise.
Results in \cite{Pola06} show that cost function $V$ is related to the probability quantity in (\ref{OptProb}) as follows:
\begin{equation}
P(\{\omega\in\Omega:x_{0}\in X_{0},x_{1}\in X_{1},...,x_{N}\in X_{N}\})=V(0,x_{0},\pi). \nonumber
\end{equation}

Hence the ODAA Problem can be reformulated, as follows:

\begin{problem}
(Optimal Dynamic Asset Allocation) Given a finite time horizon $N\in \mathbb{N}$ and the sequence of target sets in (\ref{targetsets}) compute:
\[
\pi ^{\ast }=\arg \sup_{\pi\in \Pi_{N} }V(0,x_{0},\pi ).  
\]
\end{problem}

The above formulation of the ODAA Problem is an intermediate step towards the solution of the optimal control problem under study which can now be reported hereafter.

\begin{theorem}
\label{Optimal_Theorem}
\cite{Pola06}
The optimal value of the ODAA Problem is equal to
\[
p^{\ast }=J_{0}(x_{0}),
\]
where $J_{0}(x)$ is given by the last step of the following algorithm,%
\begin{equation}
\begin{array}{ll}
J_{N}(x)=I_{X_{N}}(x), & \\
J_{k}(x)=\sup_{u_{k}\in \mathcal{U}_{k}}\int_{X_{k+1}}J_{k+1}(z)p_{f(x,u_{k},w_{k+1})}(z)dz, & k=N-1,N-2,...,1,0.%
\end{array}
\label{Bellman_eq}
\end{equation}%
\end{theorem}

The above result proposes an algorithm whose outcome is precisely the optimal Markov policy solving the ODAA Problem. A financial strategy interpretation of the above result can be found in the so--called \textit{contrarian strategy}, see e.g. \cite{Chan} and the references therein. A financial strategy in asset allocation is said to be contrarian when it suggests to buy risk asset--classes when the market is down and to sell them when the market is performing well.
In the next section we will illustrate the contrarian attitude of the proposed approach in a case study.\\
We conclude this section by briefly discussing computational issues related with the algorithm proposed in Theorem \ref{Optimal_Theorem}. Although as discussed in \cite{AbatePHDThesis}, algorithms for studying stochastic reachability for general stochastic nonlinear (and hybrid) control systems are demanding from the computational point of view, the computational complexity of the proposed algorithm remains tractable because of the following reasons: (i) the state variable of the system in (\ref{system}) is scalar;
(ii) properties of MMGM, that we use in modeling distribution of $w_{k}$, allow computational efficient solutions in solving the basic optimization problem in (\ref{Bellman_eq}) (see e.g. \cite{IAN}). Moreover we stress that the optimization process required in Theorem \ref{Optimal_Theorem} to portfolio construction can be done off--line and that design of financial products does not require tipically, severe time constraints.

\section{A Case study: A Total Return Portfolio in the US Market}\label{sec5}
In this section we illustrate the proposed methodology by designing an asset allocation in the US market.
Let us consider a total--return US fund--manager. The investment's universe consists of three asset--classes: Money--market, US Bond, and US Equity markets. Details on the indices used in the analysis are reported in the table below:
\[
\begin{tabular}
[c]{lll}
\hline
Label \hspace{5mm} &
Asset \hspace{5mm} &
Index \hspace{5mm} \\
\hline
C & Money market & US Generic T-bills 3 months\\
B & US Bond & JP Morgan US Government Bond All Maturity \\
E & US Equity & S\&P500\\
\hline
\end{tabular}
\]
Money market, bond and equity are tipically used as a portfolio ensemble in asset--allocations because of their diversified statistical properties which allow the investors to deal with different financial scenarios. \\
Time--series are in local currency (US dollars) and weekly--based from January $1$--st $1988$ to December $28$--th $2007$.
Statistical indicators\footnote{Standard deviation, skewness, kurtosis and correlation have been estimated by considering the historical average in the selected time-window daily observation. Expected return has been computed assuming a constant Sharpe--ratio \cite{luenberger} of $0.50$.} in the selected portfolio menu are reported in Table \ref{STATtab1}.
\begin{table}
\begin{center}
\begin{tabular}
{|p{3.0cm}|*{3}{c|}}
\hline &
C &
B &
E  \\
\hline
ER (ann) & 3.24\% & 5.46\% & 10.62\%  \\
\hline
SD (ann) & 0\% & 4.45\% & 14.77\%  \\
\hline
SK & 0 & -0.46 & -0.34  \\
\hline
KU & 3 & 4.25 & 5.51  \\
\hline
\end{tabular}
\begin{tabular}
{|p{2.5cm}|*{3}{c|}}
\hline 
correlation &
C &
B & 
E  \\
\hline 
C & 1 & 0 & 0  \\
\hline 
B & 0 & 1 & 0.0342   \\
\hline 
E & 0 & 0.0342 & 1   \\
\hline 
\end{tabular}
\end{center}.
\caption{Statistical indicators of the asset--classes menu.}
\label{STATtab1}
\end{table}
By comparing asset--classes ERs it is readily seen that US Equity (asset--class E) ensures higher performances than US Bond (asset--class B) which in turn, ensures higher performances than Money Market (asset--class C). On the other hand, asset--classes SDs show that US Equity is riskier than US Bond which is riskier than Money Market. 
Moreover, skewness and kurtosis reported in Table \ref{STATtab1} show that asset--classes present significant deviations--to--gaussianity (Jarque--Bera test; 99\% Confidence Level). Bond and Equity markets are leptokurtic and negative skewed;
with reference to Figure \ref{fig33}, while Asset C is in Region $5$, Assets B and E are located in Region $1$.\\
The asset allocation that we want to synthesize is characterized by a $2$--years life--time.
We consider a weekly re--balancing, i.e. we suppose to re--balance our portfolio once per week. The finite time horizon associated with this asset allocation is therefore given by $N=104$.
The investor specification is to beat a target return $7\%$ (annualized value) at maturity; his risk--budget corresponds to $7\%$ (ex--ante) monthly Value--at--Risk\footnote{Value--at--Risk (VaR) quantifies the potential loss of a portfolio to a given time horizon and to a certain confidence level. Parametric evaluation of the VaR permits to relate it to the portfolio volatility \cite{luenberger}; for example VaR to $1$ month horizon and $99\%$ confidence level is $2.3263$ SD $/\sqrt{12}$, where SD indicates the portfolio standard--deviation.} at $99\%$ Confidence--Level. \\
The above specification translates in constraints on the set $U_{k}$ of portfolio allocations and on an appropriate choice of the sequence of target sets $X_{k}$. With regard to the definition of $U_{k}$, a portfolio allocation $u_{k}\in U_{k}$ is required to satisfy budget, long-–only and risk-–budget constraints with $\sigma_{max}= $ VaR $\sqrt{12} / 2.3263=0.1042$, see footnote ($^{3}$). Moreover the target sets $X$s' formalization results in:
\begin{eqnarray}
 X_{0} &  = & \{1\},\nonumber \\
 X_{k} &  = & [0 ,+\infty[, \hspace{2mm} k=1,...,103,   \nonumber \\
 X_{104} &  = & [1.07^{2},+\infty[.
\label{sigmas}
\end{eqnarray}
The optimization criterion consists in maximizing the probability
\[
P(\{\omega\in\Omega: x_{104} \geq 1.07^2\}),
\]
where $x_{104}$ is the portfolio value at the end of the second year (investment life--time).


The first step in the methodology illustrated in the above section consists in finding a MMGM which appropriately describes the market scenario behaviour. The chosen MMGM $Y$ is composed of two multivariate gaussian random variables $Y_{1}$ and $Y_{2}$, characterized by the following univariate statistics:
\begin{eqnarray}
\mu^{1} & = &
\left[
\begin{array}
[c]{ccc}%
0.000611 & 0.001373 & 0.002340
\end{array}
\right]^{T},\nonumber\\
 \sigma^1 & = &
\left[
\begin{array}
[c]{ccc}%
0.000069 & 0.005666 & 0.019121
\end{array}
\right]^{T},\nonumber\\
\mu^{2} & = &
\left[
\begin{array}
[c]{ccc}%
0.000683 & -0.016109 & -0.017507
\end{array}
\right]^{T},\nonumber\\
 \sigma^{2} & = &
\left[
\begin{array}
[c]{ccc}%
0.000062 & 0.006168 & 0.052513
\end{array}
\right]^{T},
\label{MMGMcoeff}
\end{eqnarray}
and same correlation matrix:
\[
\begin{tabular}
{|p{0.5cm}|*{3}{c|}}
\hline 
& 
C & 
B &
E  \\
\hline 
C & 1 & 0.0633 & 0.0207  \\
\hline 
B & 0.0633 & 1 & -0.0236   \\
\hline 
E & 0.0207 & -0.0236 & 1   \\
\hline
\end{tabular}\]\\
Symbols $\mu^{i}$ and $ \sigma^{i}$ indicate (resp.) the ER and SD of the $i$--th multivariate gaussian model, which are associated with the three asset--classes. Values are weekly--based. The MMGM parameters expressing the probability related to each gaussian world are:
\begin{eqnarray}
\lambda_{1}=0.98, & \lambda_{2}=0.02.\nonumber
\end{eqnarray}
The obtained MMGM correctly represents the univariate statistics of the asset--classes up to fourth--order and the correlation patterns up to the second--order. The above MMGM has been obtained by an optimization process which minimizes the errors between
the univariate four moments associated with the time series of the asset--classes involved and the selected MMGM; the obtained error resulted in $2.24 E-7$. Additional constraints of semi--definiteness in the covariance matrix and unimodality of the univariate marginal distributions have been considered in the selection of the MMGM. 


\begin{figure}[t]
\begin{center}
\includegraphics[scale=0.8]{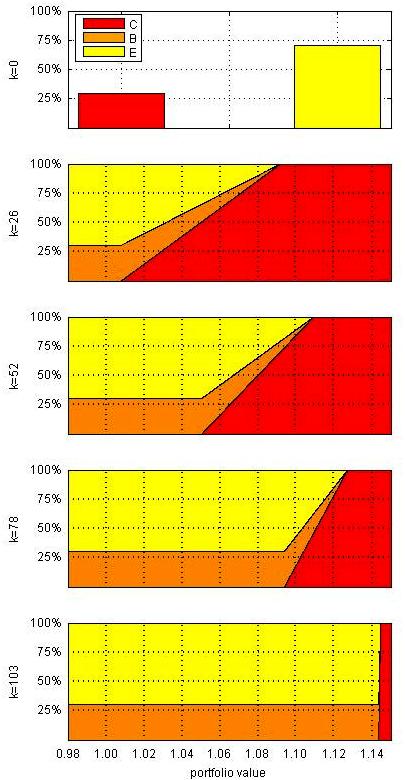}
\end{center}
\caption{Optimal portfolio allocation at times $k=1,2,...,8$.}
\label{figMAPs}
\end{figure}
By applying the methodology illustrated in the previous section the
optimal portfolio allocation is synthesized.
The optimal Markov policy at time $k=0$, corresponding to
the initial condition $x_{0}=1$, is given by
$29.50\%$ Money market, $0\%$ Bond market, and $70.50\%$ Equity market, as shown in the first panel of Figure \ref{figMAPs}. After the first week fund--manager revises the portfolio allocation. 
Optimal Markov policies corresponding to the end of each semester and 1--week before the investment maturity ($k=26, 52, 78, 103$) are reported in Figure \ref{figMAPs}. In the second panel abscissas report the portfolio value $x_{26}$ at time $k=26$. For each portfolio realization $x_{26}$, the map gives the corresponding portfolio allocation. For example a portfolio value of $x_{26}=1.000$ indicates to allocate about $30\%$ of the investors' wealth in US Bond (asset--class B) and $70\%$ in US Equity (asset--class E). It is readily seen that as the portfolio value decreases the portfolio allocation gets riskier and riskier; for example optimal policy yields $0\%$ Cash for portfolio value less than $1.0081$. This means that until this threshold is reached, the model needs to allocate the maximum allowed budget risk expressed by the Value at Risk $99\%$ constraint. In the other way around as performances gets better and better the model suggests to reduce the risk exposure: for a portfolio value greater than $1.0919$, a $100\%$ Cash allocation guarantees to reach the performance target. The attitude of this optimal strategy to increase risk exposure in presence of portfolio draw--downs, and conversely to reduce it in case of positive performance is known in the literature as \textit{contrarian strategy} \cite{Chan}. 
Optimal Markov policies at times $k=52, 78$ exhibit similar characteristics to the one for $k=26$, as shown in Figure \ref{figMAPs}. It is readily seen that as $k$ increases the portfolio re--balancing gets sharper and sharper: as time passes by, portfolio allocation strategy becomes more and more aggressive and indicates to buy more and more US Equity (asset--classes E) in the attempt to reach the target goal.
%
Main differences 
is that the optimal Markov policy require to be riskier than the one corresponding to the previous map. Let us consider a portfolio value equal to 1.06. At $k=26$ the optimal allocation is $62\%$ Cash, $12\%$ Govy, $26\%$ Equity. After a semester ($k=52$), the optimal allocation for the same portfolio value is $16\%$ Cash, $26\%$ Govy, $58\%$ Equity. At $k=78$ the optimal allocation is $0\%$ Cash, $30\%$ Govy, $70\%$ Equity. The last map at $k=103$ presents similar feature. It should be stressed that the optimal solution is particularly sharp. For a portoflio value equal to $1,1435$ the model suggests to use all the available budget risk, while for a portfolio value equal to $1,1446$ a full Cash allocation ensures to reach the target.

\begin{figure}[t]
\begin{center}
\includegraphics[scale=0.25]{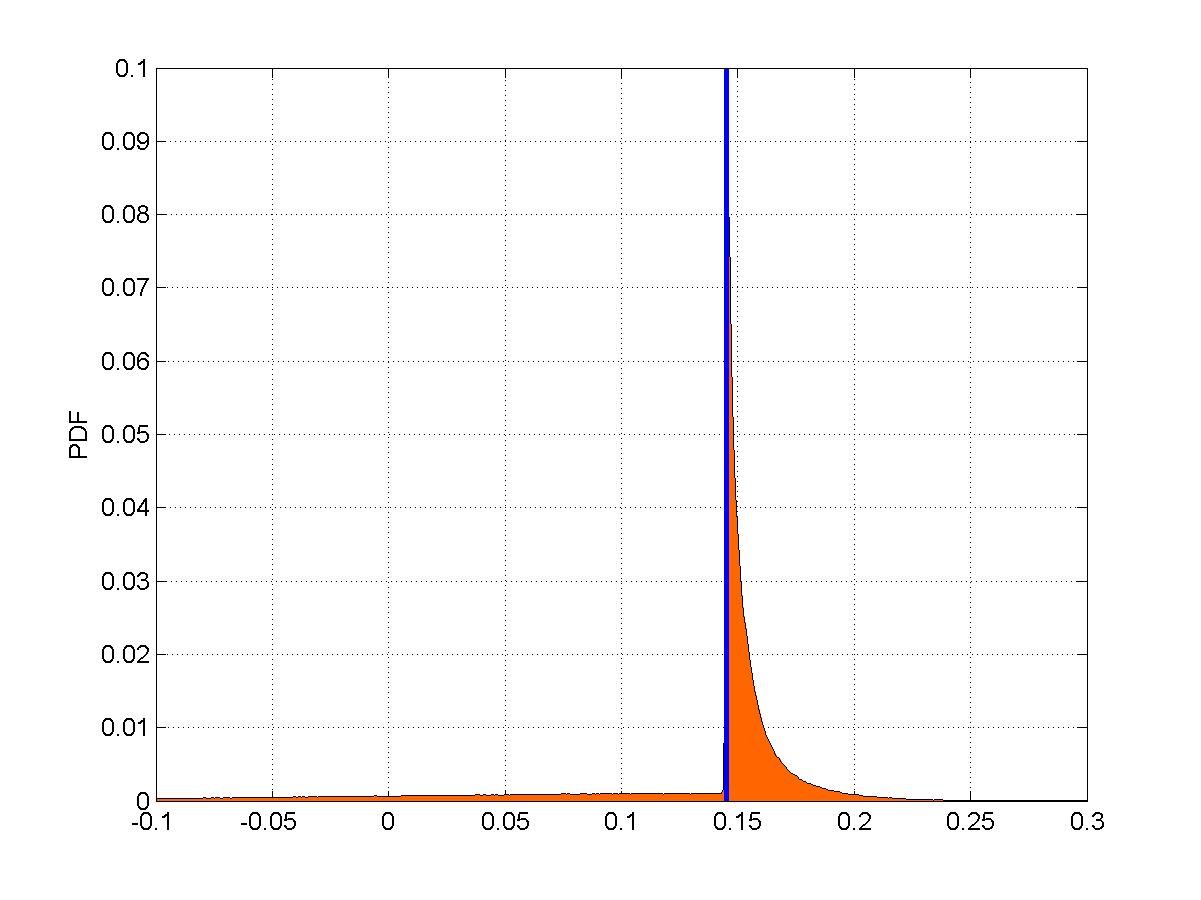}
\end{center}
\caption{Return Localization.}
\label{fig11}
\end{figure}

The maximal probability of achieving the investment goal is:
\[
p^{\ast}=J_{0}(x_{0})=77.62\%.
\]
In order to make a validation of this result we run a Montecarlo simulation with $10^{6}$ scenarios. 
The results yield a probability of $77.76\%$ and are depicted in Figure \ref{fig11}. The different probability values achieved by the optimization process and the Montecarlo simulations are due to the numerical errors in the optimization process and to the number of realizations chosen in the Montecarlo simulations; nevertheless the two probability values are close one each other. Figure \ref{fig11} shows that most of the realizations are placed at the right of the portfolio value $\underbar{x}=1.07^{2}$ as required by the target sets specifications in (\ref{targetsets}).

In order to make a comparison with traditional standard approaches in asset--allocation, we run the same exercise for a Markowitz investor with a finite time horizon $N=104$. In this case the optimal solution requires to invest the full budget--risk, corresponding to an allocation of
$0\%$ Cash, $30\%$ Govy, $70\%$ Equity. This allocation has been derived at first by optimizing according to the mean--variance paradigm, and then by selecting the optimal portfolio on the efficient frontier which maximizes the probability to beat the target return in two years horizon. The resulting probability of achieving the investment target goal is $61.90\%$. This probability has been evaluated by performing a Montecarlo simulation with $10^6$ scenarios. 
By comparing the probability of achieving the target goal in the two approaches it is readily seen that the ODAA approach provides a differential of $15.86\%$ with respect to the Markowitz approach, which is remarkable in portfolio allocation in finance industry.

\section{Conclusion}\label{sec6}

In this paper we addressed Optimal Dynamic Asset Allocation. Given a
specified finite time horizon and a sequence of target sets that the investors
would like their portfolio to stay within, the optimal portfolio allocation is
synthesized in order to maximize the joint probability for the portfolio value
to fulfill the target sets requirements. The proposed approach has been shown to overcome asset--classes gaussian--based modeling and static portfolio allocation limits of the methodologies currently in use in finance industry. \\
Contrarian strategy attitude of the proposed methodology can result in being rather aggressive if applied to some specific financial products 
in case of large market draw downs. The authors are currently investigating some approaches based on stochastic hybrid systems modeling \cite{BujorianuHSCC04} with the goal of making the proposed methodology suitable of application to a wider range of financial products.

\bigskip
\textbf{Aknowledgement.} The authors would like to thank Maria Domenica Di Benedetto (University of L'Aquila, Center of Excellence DEWS, Italy), John Lygeros (ETH, Switzerland) and Roberto Dopudi (Cr\'edit Agricole Asset Management SGR, Italy) for stimulating discussions on the topic of this paper.

\section*{Appendix: time series--providers}
Data--providers and details of the time--series listed in Figure \ref{fig12} are:
\begin{itemize}
\item FX rates. Bloomberg;
\item Government Bonds. JP Morgan (All Maturity indices);
\item Inflation--Linked Bonds. Barclays Capital International (All Maturity and Total Return indices);
\item Corporate Bonds and High Yields. Merril Lynch;
\item Equity markets. Morgan Stanley Capital International;
\item Convertible. UBS (Total Return and At-The-Money indices);
\item Real Estate. EPRA/NAREIT (Total Return and At-The-Money indices);
\item Commodities. Goldman Sachs (Total Return and At-The-Money indices).
\end{itemize}
All time--series are in local currencies, Commodities are in USD. Bloomberg \cite{BLOOMBERG} code of the asset--classes
listed in Figure \ref{fig12} are (resp.):

\begin{table*}[htbp]
\begin{center}
\begin{tabular}{llllll}
gbpeur curncy, &	jpyeur curncy, & usdeur curncy,	& cadeur curncy,	& audeur curncy, \\
jpmgemlc index, & jpmtus index, &	jnycjp index, &	bcee1t index,	& bcit1t index,	\\
bciu1t index,	& c0a0 index, & er00 index, &	j0a0 index,	& he00 index, \\
msdlemu index, &	msdlus index, &	msdluk index,	& msdljn index,	& msdlhk index,	\\
msdlsg index, &	mselegfm index, &	mselegfl index,	& mselegfa index, & mseltcf index, \\
mxemsc index,	& mcldus index,	& mclduk index, &	mclajn index,	& mclahk index,	\\
mclasg index,	& ucbiemae index, &	ucbijvaj index,	& ucbiusam index,	& exuk index, \\
ugna index, & eljp index,	& gscitrsi index. & &
\end{tabular}
\end{center}
\end{table*}

\bibliographystyle{alpha}
\bibliography{biblio1}

\begin{thebibliography}{APLS08}

\bibitem[Aba07]{AbatePHDThesis}
Alessandro Abate.
\newblock {\em Probabilistic Reachability for Stochastic Hybrid Systems:
  Theory, Computations, and Applications}.
\newblock PhD thesis, EECS Department, University of California, Berkeley, Nov
  2007.
\newblock Available at \texttt{http://www.eecs.berkeley.edu/$\sim$aabate}.

\bibitem[APLS08]{AbateAut08}
A.~Abate, M.~Prandini, J.~Lygeros, and S.~Sastry.
\newblock Probabilistic reachability and safety for controlled discrete time
  stochastic hybrid systems.
\newblock {\em Automatica}, 44(11):2724--2734, 2008.

\bibitem[Bar00]{BAR}
N.~Barberis.
\newblock Investing for the long run when returns are predictable.
\newblock {\em The Journal of Finance}, 55:225--264, 2000.

\bibitem[Ber01]{bertsekas}
{D.P.} Bertsekas.
\newblock {\em Dynamic Programming and Optimal Control}.
\newblock Athena Scientific, Belmont Massachusetts, 2001.

\bibitem[BL03]{BujHSCC03}
{M.L.} Bujorianu and J.~Lygeros.
\newblock Reachability questions in piecewise deterministic markov processes.
\newblock In O.~Maler and A.~Pnueli, editors, {\em Hybrid Systems: Computation
  and Control}, volume 2623 of {\em Lecture Notes in Computer Science}, pages
  126--140. Springer Verlag, Berlin, 2003.

\bibitem[BLO]{BLOOMBERG}
Bloomberg.
\newblock http://www.bloomberg.com.

\bibitem[BSL97]{BRE}
{M.J.} Brennan, {E.S.} Schwartz, and R.~Lagnado.
\newblock Strategic asset allocation.
\newblock {\em Journal of Economic Dynamics and Control}, 21:1377--1403, 1997.

\bibitem[BSS08]{IAN}
I.~Buckley, D.~Saunders, and L.~Seco.
\newblock Portfolio optimization when asset returns have the gaussian mixture
  distribution.
\newblock {\em European Journal of Operational Research}, 185:1434–--1461,
  2008.

\bibitem[Buj04]{BujorianuHSCC04}
M.~L. Bujorianu.
\newblock Extended stochastic hybrid systems and their reachability problem.
\newblock In R.~Alur and George~J. Pappas, editors, {\em Hybrid Systems:
  Computation and Control 2004}, volume 2993 of {\em Lecture Notes in Computer
  Science}. Springer-Verlag, Philadelphia, PA, USA, 2004.

\bibitem[BX02]{BREXIA}
{M.J.} Brennan and Y.~Xia.
\newblock Dynamic asset allocation under inflation.
\newblock {\em The Journal of Finance}, 57(3):1201--1238, 2002.

\bibitem[Cha88]{Chan}
K.~C. Chan.
\newblock On the contrarian investment strategy.
\newblock {\em Journal of Business}, 61:147–--163, 1988.

\bibitem[Lue98]{luenberger}
{D.G.} Luenberger.
\newblock {\em Investment Science}.
\newblock Oxford University Press, New York, USA, 1998.

\bibitem[Mar52]{markowitz}
H.~Markowitz.
\newblock Portfolio selection.
\newblock {\em The Journal of Finance}, 7(1):77--91, 1952.

\bibitem[Mer69]{MER}
{R.C.} Merton.
\newblock Lifetime portfolio selection under uncertainty: the continuous--time
  case.
\newblock {\em Review of Economics and Statistics}, 51:247--257, 1969.

\bibitem[PLB06]{Pola06}
G.~Pola, J.~Lygeros, and {M.D.}~Di Benedetto.
\newblock Invariance in stochastic dynamical systems.
\newblock In {\em 17--th International Symposium on Mathematical Theory of
  Network and Systems}, Kyoto, Japan, July 24th -- 28th 2006.

\bibitem[PP06]{PolaCDC06}
G.~Pola and G.~Pola.
\newblock Optimal dynamic asset allocation: A stochastic invariance approach.
\newblock In {\em 45th IEEE Conference on Decision and Control}, pages
  2589--2594, San Diego, CA, December 2006.

\bibitem[Sam69]{SAM}
{P.A.} Samuelson.
\newblock Lifetime portfolio selection by dynamic stochastic programming.
\newblock {\em Review of Economics and Statistics}, 51:239--246, 1969.

\bibitem[Xia01]{XIA}
Y.~Xia.
\newblock Learning about predictability: the effect of parameter uncertainty on
  dynamic asset allocation.
\newblock {\em The Journal of Finance}, 56(1):205--246, 2001.

\end{thebibliography}

\end{document}